\pdfoutput=1

\documentclass[
    Template=acmconf,
    PublicVersion=true,
    SubmissionID=3,
]{config/paper}

\acmConference{3rd Workshop on Heterogeneous Composable and Disaggregated Systems}{Apr 28, 2024}{San Diego, CA, USA}

\newcommand{\natitle}{
    \title{Fork is All You Need in Heterogeneous Systems}

    \author{Zixuan Wang}
    \email{zxwang@ucsd.edu}
    \affiliation{
        \institution{University of California, San Diego}
        \country{}
    }
    \author{Jishen Zhao}
    \email{jzhao@ucsd.edu}
    \affiliation{
        \institution{University of California, San Diego}
        \country{}
    }

    \maketitle
}

\begin{document}

\napaperpagestyle{}

\begin{abstract}

We present a unified programming model for heterogeneous computing systems.
Such systems integrate multiple computing accelerators and memory units to deliver higher performance than CPU-centric systems.
Although heterogeneous systems have been adopted by modern workloads such as machine learning, programming remains a critical limiting factor.
Conventional heterogeneous programming techniques either impose heavy modifications to the code base or require rewriting the program in a different language.
%
% This not only limits the conventional workloads to utilize the heterogeneous systems but also holds back the adoption of emerging architectures and memory in existing heterogeneous systems.
%
Such programming complexity stems from the lack of a unified abstraction layer for computing and data exchange, which forces each programming model to define its abstractions.
However, with the emerging cache-coherent interconnections such as Compute Express Link, we see an opportunity to standardize such architecture heterogeneity and provide a unified programming model.
We present CodeFlow, a language runtime system for heterogeneous computing.
CodeFlow abstracts architecture computation in programming language runtime and utilizes CXL as a unified data exchange protocol.
%
% \zixuan{Maybe not mention WebAssembly here to reduce confusion, given that the abstract is fairly short and people may not fully understand why it needs WebAssembly.}
%
Workloads written in high-level languages such as C++ and Rust can be compiled to CodeFlow, which schedules different parts of the workload to suitable accelerators without requiring the developer to implement code or call APIs for specific accelerators.
CodeFlow reduces programmers' effort in utilizing heterogeneous systems and improves workload performance.
\end{abstract}

\natitle{}

\section{Introduction}

% - CXL programming
%     - Current problems in PCIe accelerator programming
%         - It’s mostly RPC-style: host copies data to accelerator, copies code, gives a signal, and collects results
%         - Not a clean or unified programming
%     - What already works in existing cache-coherent system
%         - CPU to CPU cache-coherence, enables using `fork`
%         - Fork a thread, do computation on remote NUMA CPU node, collects results
%         - Most, if not every, programmers know the concept
%     - Now with CXL, we can think about offloading computation to CXL device, using `fork` styled computation
%         - The same C code

%         ```c
%         do_computation_host
%         fork {
%         	do_computation <-- can be scheduled on host CPU, or CXL device
%         }
%         get_fork_results
%         ```

%         - It’s a new era of heterogeneous programming
%         - Can write a quick proposal (extended abstract) and discuss with Geoff
%     - Title: “Fork is all you needed”

\overview{Modern workloads heavily rely on heterogeneous systems.}
Such systems integrate multiple types of specialized computing and memory devices toward high-performance computing.
GPU and high-bandwidth memory (HBM) support high-performance machine learning systems, SmartNICs accelerate network processing in cloud computing, and processing-in-memory (PIM) improves computing efficiency in memory-intensive workloads.
With the advancement of device interconnection technologies such as PCIe, NVLink, and Compute Express Link (CXL), more types of accelerators and memory will be integrated into heterogeneous systems to support irregular performance requirements of workloads such as large language models, graph processing, and serverless.

\overview{Software support remains one critical limitation of heterogeneous systems' scalability.}
Historically, accelerators have established unique protocols to exchange data and manage tasks, relying on specialized operating system drivers and programming libraries.
Thus, to use heterogeneous systems, applications must employ specialized code for accelerators while ensuring compatibility between them, and the operating system must manage different drivers.
Such programming complexity drives the development of programming supports for heterogeneous computing.
However, they either introduce unconventional programming models~\cite{sycl, opencl, oneapi} requiring rewriting applications, or targeting domain specific use cases such as machine learning~\cite{pytorch, TornadoVM,triton-lang} making them hard to generalize.

\begin{figure}[t!]
    \centering
    \nafig{B}{\linewidth}
        {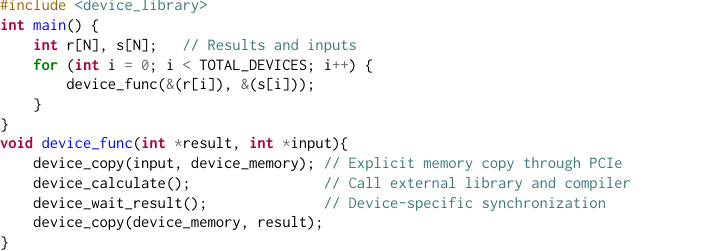}
        {subfig:hetero-code-example}
        {Heterogeneous code example.}
    \\
    \nafig{B}{\linewidth}
        {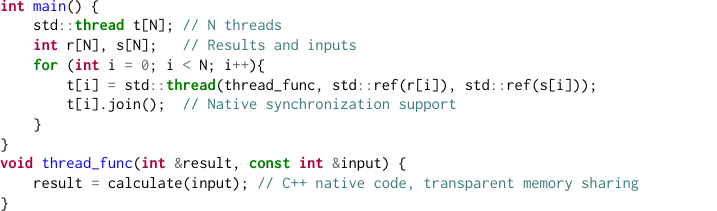}
        {subfig:multi-thread-code-example}
        {Multithreading code example.}
    % \xsubfigure{A}{
    %     width=.4\linewidth,
    %     body={\includegraphics[width=\linewidth]{figure/fig/tex/hetero-code-example.pdf}},
    %     caption={\label{subfig:hetero-code-example}Heterogeneous code example.}
    % }
    % \xsubfigure{B}{
    %     width=.4\linewidth,
    %     body={\includegraphics[width=\linewidth]{figure/fig/tex/multi-thread-code-example-cpp.pdf}},
    %     caption={\label{subfig:multi-thread-code-example}Multithreading code example.}
    % }
    % \makerow{A,B}
    \caption{Heterogeneous programming is generally more complex than multi-threading programming due to the need of using external libraries, system drivers, and compiler toolchains.}
    \label{fig:hetero-vs-multithread}
\end{figure}

\overview{Programming heterogeneous systems is a special type of parallel programming where CPUs send data and code to accelerators, signal them to start tasks, and wait for results.}
%
% Similarity between conventional heterogeneous computing and multi-threading computing. Point out why conventional heterogeneous computing not able to generalize due to cache coherence
%
% It follows a remote procedure call (RPC) styled programming model where CPUs send data and code to accelerators, signal them to start tasks, and wait for results.
%
In this process, CPUs and accelerators work in parallel and communicate through PCIe or similar interconnection buses.
%
% \updatelater{This RPC-styled model introduces clutters, give an example. However multithreading computing is much more intuitive, give an example, then point out difference is cache coherence}
%
Such parallel computing in heterogeneous systems (\figref{subfig:hetero-code-example}) is conceptually similar to multi-CPU systems (\figref{subfig:multi-thread-code-example}), where multiple CPU cores run different program threads and communicate through core-to-core buses.
However, multi-CPU programming has a much more intuitive programming model where the parallel tasks are defined as native functions and spawned by \texttt{fork} or \texttt{clone} system calls which are natively supported by commonly used languages such as C, C++, Rust.
Such multi-threading models are enabled by (1) the unified CPU architecture that enables the code to be implemented in a unified language, and (2) the cache-coherent connections between CPUs that enables different threads to share memory transparently without relying on explicit calls to libraries.
Historically, these two architectural properties are not available in heterogeneous systems, because different accelerators implement their unique architectures, and they are typically interconnected through non-cache-coherent peripheral buses such as PCIe.
With the recent advancement of Compute Express Link (CXL) as a general cache-coherent interconnection, we see an opportunity to generalize heterogeneous computing.

\zixuan{Our goal is to ...}
\overview{We present CodeFlow, a unified programming model for heterogeneous computing.}
%
% By leveraging language runtime system to abstract architectures and employing CXL to coherently connect accelerators, we present a unified programming model that runs unmodified multi-threading code in heterogeneous systems.
%
CodeFlow leverages WebAssembly System Interface (WASI)~\cite{wasi} to build a language runtime system, which serves as an intermediate layer between higher-level language and lower-level heterogeneous architectures.
This runtime system allows programs to be implemented in a single language and compiled to a single runtime representation, without the need to explicitly implement code for different architectures and compile with multiple toolchains.
At low level, CodeFlow leverages CXL to enable coherent memory sharing among heterogeneous accelerators, thus minimizing explicit library calls for cross-devices data movements.

\overview{We demonstrate that CodeFlow can significantly simplify heterogeneous programming.}
The heterogeneous code can now be written as plain multi-threading code, which leverages the language's native supports for standard multi-threading mechanisms, such as synchronization and memory sharing.
%
% Such code is then compiled to a unified language runtime representation which is based on .
%
The runtime schedules different threads to run on different accelerators, handles memory sharing through CXL, and just-in-time compiles code for accelerator architectures.
CodeFlow also allows conventional multi-threading programs--designed for multi-CPU systems only--to leverage heterogeneous systems by re-compiling with CodeFlow toolchain.

\section{Background and Motivation}

Heterogeneous computing is similar to multi-threading in many ways.
We observe a few fundamental limitations in existing heterogeneous computing that hold back the system scalability and point out potential solutions with the recent advancements in programming and architecture techniques.

\begin{figure}[t]
    \centering
    \includegraphics[width=.9\linewidth]{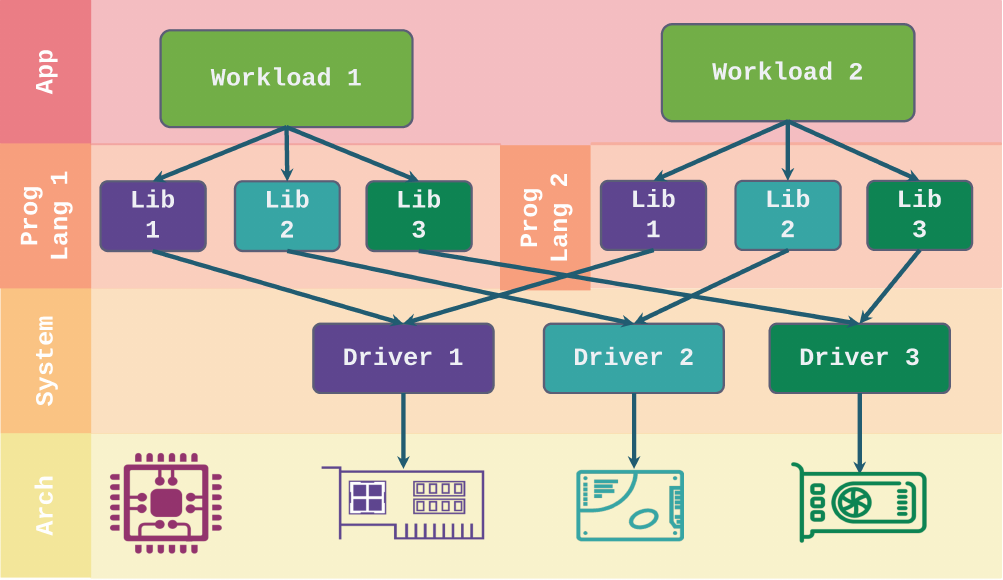}
    % \nafig{b}{\linewidth}{figure/fig/hetero-system-stack.pdf}{subfig:hetero-system}{Heterogeneous System Stack.}
    % \nafig{b}{.48\linewidth}{example-image-a}{subfig:hetero-prog-example}{Code Example.}
    \caption{Heterogeneous system stack with conventional programming techniques, where workloads have to use multiple libraries and compilers to utilize underlying heterogeneous architectures. Such programming complexity limits the system's scalability of adding new architectures.}
    \label{fig:hetero-system-stack}
\end{figure}

\subsection{Multi-Threading Programming}

\overview{Multi-threading is a programming approach that allows a process to be divided into multiple threads, with each thread capable of executing concurrently.}
This model enables applications to perform multiple operations simultaneously, leveraging the computational power of modern multi-CPU systems more effectively.
In such systems, cache coherence plays a critical role at architecture level, ensuring the consistency of data cross the caches of a multi-core or multi-processor systems.
When multiple threads running on different CPU cores access shared data, cache coherence mechanisms ensure that all modifications are propagated across all caches.
This guarantees that any thread accessing a shared data will see the most recent update, irrespective of which core the updating thread runs on.

Cache coherence fundamentally simplifies the multi-threading programming model while ensuring correctness and improving performance.
Cache coherence allows developers to write multithreaded applications without the cumbersome need to manually synchronize data between threads or processors.
%
% This abstraction not only reduces the complexity of coding for concurrency but also helps applications to scale efficiently with the addition of more CPU cores.
This abstraction reduces the complexity of coding for concurrency and helps applications scale efficiently by adding more CPU cores.

\subsection{Heterogeneous Systems}

\overview{Heterogeneous systems integrate diverse processors and accelerators, such as CPUs, GPUs, FPGAs, and DSPs, to leverage their specialized computational capabilities for enhanced performance and efficiency across various applications.}
This integration, while beneficial for addressing the irregular demands of modern computing tasks such as machine learning and graph processing, introduces significant programming challenges.
As shown in \figref{fig:hetero-system-stack}, developers are imposed to utilize multiple programming models and languages, alongside the use of specialized libraries designated to each architecture.
%
% Additionally, the orchestration of code and optimization across these diverse architectures demands a broad skill set and a deep understanding of each component's unique capabilities and limitations.
%
Although many libraries and languages emerges to serve as an intermediate layer between developers and heterogeneous systems, aiming to provide easy-to-use programming interfaces, they either introduce new programming models~\cite{sycl, opencl} or target scoped use cases, requiring application rewriting and raising the complexity of adopting new architectures.

Such programming complexity in heterogeneous systems is mainly caused by (1) the absence of cache coherence and (2) the diversity of architectures.
The lack of a unified cache coherence mechanism leads to data inconsistencies, requiring explicit data management and synchronization efforts.
Moreover, the need to navigate and optimize for different execution models and memory systems adds to the development burden.
These two challenges underscore the need for unified programming models and toolchains that can simplify data management and exploit the full potential of heterogeneous systems.
In following sections, we describe two key techniques that enable us to address the cache coherence and architecture heterogeneity challenges respectively.

\begin{figure*}[t]
    \centering
    \includegraphics[width=\linewidth]{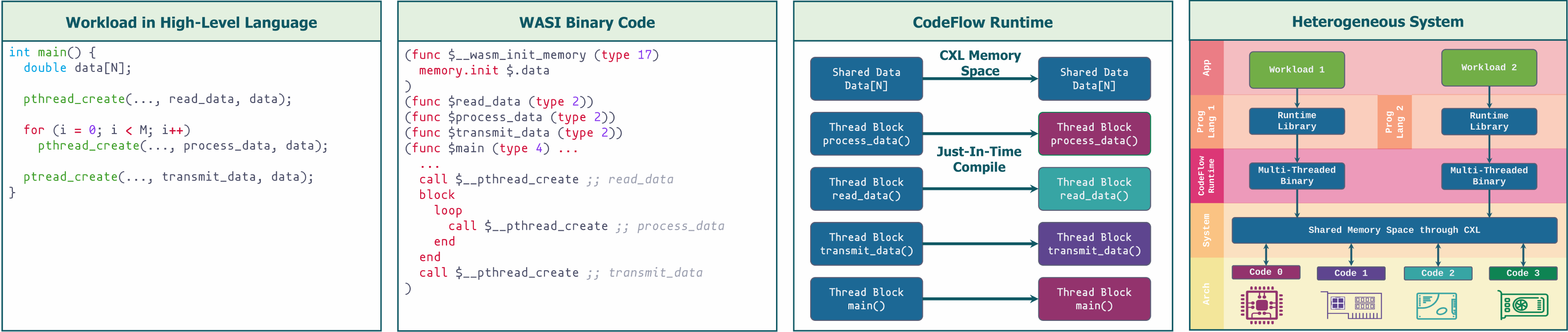}
    \caption{Heterogeneous system programming flow with CodeFlow.}
    \label{fig:codeflow-system-stack}
\end{figure*}

\subsection{Compute Express Link}

Compute Express Link (CXL)~\cite{cxl} is an open industry-standard interconnection protocol, designed to facilitate high-speed communications between processors, accelerators (such as GPUs and SmartNICs), and memory devices in computing systems.
A unique feature of CXL is its capability at hardware level to enable coherent memory access across heterogeneous architectures.
This coherence ensures that various processors and accelerators can transparently share memory resources, as though they were part of a unified system.
By leveraging CXL, we can simplify the memory access operations in heterogeneous programming models.

\subsection{WebAssembly System Interface}

WebAssembly is a language runtime system initially designed to accelerate web application.
%
% It defines a hardware-agnostic instruction set which allows high-level languages (such as JavaScript, C, C++, Rust) to be compiled as WebAssembly and run on browsers on any hardware.
It is a hardware-agnostic binary instruction format designed as a portable compilation target for high-level languages like C, C++, and Rust.
It enables code to run with near-native performance across various platforms, initially within web browsers to improve the performance of web applications.
%
% Recently, WebAssembly has been extended to run beyond web browser into general computing environments such as bare metal Linux systems, by leveraging WebAssembly System Interface (WASI)~\cite{wasi}.
%
Recently, WebAssembly System Interface (WASI) has been proposed to provide WebAssembly applications with a standardized system interface, allowing them to interact with the operating system's underlying capabilities, such as file systems, network sockets, and other system resources.
With WASI, high-level code can now compile to WebAssembly, run in WASI environment, access system resources through standardized interface, and execute on any suported architecture, all without rewriting or recompiling the code.
This portability makes WASI a perfect candidate for us to design an abstraction layer between high-level code and low-level heterogeneous architectures.

% \updatelater{
% WebAssembly System Interface (WASI) is a modular system
% interface for WebAssembly, a low-level binary format for executing
% code in a web browser or other environments. It provides a standard
% set of APIs that enable WebAssembly programs to run outside the
% web browser in a standalone environments, such as servers, desktop
% applications, and embedded devices.

% By providing a standard system interface, WASI helps to ensure
% compatibility and portability of WebAssembly programs across dif-
% ferent environments, and helps to eliminate the need for developers
% to write platform-specific code. Additionally, WASI provides a se-
% cure sandbox for executing WebAssembly code, which helps to
% mitigate security risks and prevent malicious code from accessing
% sensitive system resources.
% }

\section{Design}

This section introduces the design of CodeFlow framework, from compilation, runtime management, to CXL-based shared memory and heterogeneous system mapping.
% %
% \figref{fig:codeflow-system-stack} shows the overview of programming flow in CodeFlow systems.
%

\subsection{Compile High-Level Code}

As shown in \figref{fig:codeflow-system-stack}, the workload code is written in high-level language as a multithreading program.
The workload is divided as a few sub-tasks each running as a standalone thread, sharing the access to the shared memory.
In this example, each thread is responsible for a specific job, from reading the data from the file, to processing the data in parallel, and finally transmitting the data.
All threads are implemented in native language (C in this example) without using any hardware-specific library or compiler.
The program invokes the C standard libraries provided by WASI~\cite{wasi-libc}, which implement native library functionalities while inherently interact with the WebAssembly runtime system instead of the bare metal operating system.
The workload can call third-party libraries for specific tasks such as network I/O, and all libraries are compiled to WASI binary format with the workload.

WASI defines a hardware-agnostic assembly binary representation by extending the WebAssembly binary representation with capabilities to access operating system resources such as filesystem.
The high level workload is compiled once as WASI binary, where each thread is represented by a WASI code block, as shown in \figref{fig:codeflow-system-stack}.
Such WASI binary is then executed by CodeFlow runtime system which schedules threads in the heterogeneous system.

\subsection{Runtime System}

CodeFlow runtime system executes WASI multi-threading binaries in the CXL-interconnected heterogeneous system.
On loading a WASI binary, CodeFlow analyzes each thread's code and detects a processor or accelerator to execute the thread.
E.g., a file access thread will be assigned to CPU or an in-storage processor, a parallel data processing thread will be scheduled to use GPU, and network traffic is scheduled to use SmartNICs.
Developers can annotate the workload source code to assist CodeFlow's device detection process.
Once devices are enumerated, CodeFlow performs just-in-time (JIT) compilation that compiles the WASI binary to the device's code and executes it.
This code generation is built on-top-of conventional heterogeneous computing libraries, including device libraries such as CUDA, and frameworks such as PyTorch~\cite{pytorch} and SYCL~\cite{sycl}.
The workload's shared memory is placed in the CXL memory space shared by processors and accelerators to provide coherent access across threads.
To minimize the JIT overhead, the developer can choose to perform ahead-of-time (AOT) compilation that compiles the WASI binary to device code ahead-of-time, thus reducing the JIT time consumption at runtime.

\subsection{CXL Interconnection}

The heterogeneous system is interconnected through CXL, where processors and accelerators share a coherent memory address space, allowing all connected devices to coherently access such memory space.
Following CXL's specification, an accelerator with device memory is defined as a CXL Type 2 device, and a pure memory device is defined as a CXL Type 3 device.
The heterogeneous system could be composed by a combination of Type 2 and 3 devices and their on-device memory has divergent performance characteristics.
CodeFlow places shared data in CXL memory space for coherent accesses from different devices.
To maximize the workload performance, CodeFlow plans the data placement in device memory (e.g., GPU memory) as part of its profiling-guided optimizations at runtime, and such data can be migrated between devices to maximize temporal access performance~\cite{tpp}.

\ctable[
    caption = {System for Evaluation},
    label = {subfig:system-configuration},
    width = \linewidth{},
    doinside = \small,
    pos = t
]{lX<{\centering}}{
    % No table footnote
}{
                                                                             \FL
    \multirow{2}{*}{CPU}            & Intel SapphireRapids 4416+             \NN
    % ~                               & 950 KiB L1D\$, 40 MiB L2\$, 37.5 L3\$  \NN
    ~                               & 2 Sockets, 20 cores per socket         \ML
    Memory                          & 256 GiB DDR5                           \ML
    \multirow{3}{*}{CXL}            & Intel Agilex-7 FPGA                    \NN
    ~                               & CXL hard IP type 2 \& type 3           \NN
    ~                               & 64 GiB DDR4 SODIMM                     \ML
    OS                              & Ubuntu 22.04, Linux 6.7                \LL
}

% Line endings \FL \ML \LL, and empty line \NN

% L1d:                   1.9 MiB (40 instances)
% L1i:                   1.3 MiB (40 instances)
% L2:                    80 MiB (40 instances)
% L3:                    75 MiB (2 instances)

\ctable[
    caption = {CXL Performance},
    label = {table:cxl-perf},
    width = \linewidth{},
    doinside = \small,
    pos = t
]{lX<{\centering}X<{\centering}}{
    % No table footnote
}{
                                                   \FL
    ~           & \multicolumn{2}{c}{Performance}  \tabularnewline \cmidrule{2-3}
    Component   & Latency (ns) & Bandwidth (GiB/s) \ML
    Local DDR5  &        108.2 &             105.0 \NN
    Remote DDR5 &        171.5 &              59.1 \NN
    Local CXL   &        371.2 &              17.4 \NN
    Remote CXL  &        538.0 &               9.0 \LL
}

% Line endings \FL \ML \LL, and empty line \NN

% L1d:                   1.9 MiB (40 instances)
% L1i:                   1.3 MiB (40 instances)
% L2:                    80 MiB (40 instances)
% L3:                    75 MiB (2 instances)

\section{Evaluation}

We build an experimental system with CXL-capable CPUs and FPGAs.
As shown in \tabref{subfig:system-configuration}, we use a two-socket Intel CPU machine.
We configured the BIOS to enable CXL connections between CPU and devices on PCIe bus.
We implemented CXL devices in Intel Agilex-7 FPGAs by invoking Intel's CXL hard IP on the FPGA, connecting these IPs with on-board memory, and presenting the FPGA as a CXL device to the system.
We modified the Linux kernel's CXL driver to recognize our CXL devices due to the incompatibilities between our CXL 1.1/2.0 IPs and the Linux kernel's implementation.
We implemented CodeFlow on top of Wasmtime~\cite{wasmtime} to be compatible with WASI SDK 21 with thread extension, and evaluated the system's performance.

\subsection{CXL Performance}

% We evaluated our CXL device performance with LENS~\cite{}

We built a memory performance microbenchmark suite based on prior research on heterogeneous memory systems~\cite{nvleak,LENS-VANS}, and use it to profile the CXL device performance.
This benchmark suite uses strided memory access with pointer-chasing pattern to measure memory access latency.
\tabref{table:cxl-perf} shows our evaluation of latency and bandwidth of CXL compared to CPU system memory.
The \emph{Local} represents performance between a local NUMA CPU and a local NUMA/CXL memory and \emph{Remote} represents performance between remote NUMA nodes.
We find that local CXL latency is much lower than remote CXL latency, while being higher than CPU system memory.
The CXL memory bandwidth is lower than remote CPU memory bandwidth.
This is because our Intel FPGA has DDR4 on-board memory which is slower than DDR5, and FPGA CXL controller adds up latencies due to FPGA frequency limited to 475MHz.
We expect production CXL memory to have higher performance.
\zixuan{Compare CXL latency with PCIe latency}

\begin{figure}[t]
    \centering
    % % \nafig{b}{.48\linewidth}{example-image-a}{subfig:wasi-microbench}{WASI microbenchmark.}
    % % \nafig{b}{.48\linewidth}{example-image-a}{subfig:wasi-app-performance}{Workload performance.}
    % \begin{subfigure}[b]{.75\linewidth}
    %     \centering
    %     % \includegraphics[width=\linewidth, trim={3.5cm 10.1cm 4cm 11.28cm}, clip=true]{figure/fig/external/wasi-llc-miss.pdf}
    %     \includegraphics[width=\linewidth]{figure/plot/wasi-llc-miss.tikz.pdf}
    %     \caption{CPU LLC miss ratio in random read.}
    %     \label{subfig:wasi-llc-miss}
    % \end{subfigure}
    % \\
    % \begin{subfigure}[b]{.75\linewidth}
    %     \centering
        % \includegraphics[width=\linewidth, trim={3.5cm 10.2cm 4cm 11.1cm}, clip=true]{figure/fig/external/wasi-ptr-chasing.pdf}
        \includegraphics[width=\linewidth]{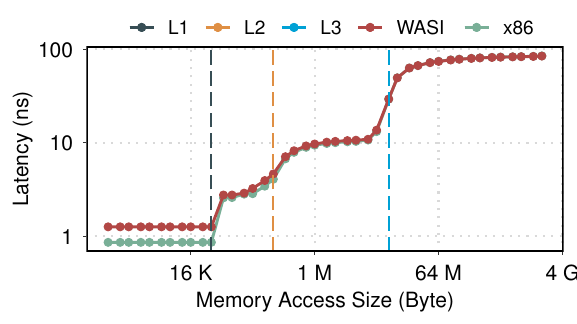}
    %     \caption{Memory access latency in pointer-chasing read.}
    %     \label{subfig:wasi-ptr-chasing}
    % \end{subfigure}

    \caption{CodeFlow runtime performance compared to running native x86 code, using pointer-chasing read.}
    \label{fig:wasi-perf}
    \vspace{-10pt}
\end{figure}

\subsection{WASI Performance}

We adapt our microbenchmark suite to run in WebAssembly and profile the CodeFlow runtime system performance.
% %
% \figref{subfig:wasi-llc-miss} shows the CPU last-level-cache (LLC) load miss rate when executing random memory access for CodeFlow's WASI code and x86's native code.
% %
% The result shows that CodeFlow experiences a lower LLC load miss, which is due to the runtime systems' optimizations when accessing memory.
% %
% But once the memory footprint reaches beyond the LLC size, CodeFlow has the same miss rate as x86 native code, showing that CodeFlow handles memory-intensive workloads with near-native performance.
%
\figref{fig:wasi-perf} shows our evaluation of memory access latency given different memory access sizes.
We use a pointer chasing memory access pattern to minimize the system noises (from cache prefetchers and compiler optimizations) in latency measurements.
We observe that CodeFlow has the same latency compared to x86 except a slightly higher latency when memory region is with CPU L1 cache size.
This shows that CodeFlow introduces very minimal memory footprint overhead when accessing memory.

\section{Future Directions}

We propose a new programming model in heterogeneous systems allowing workload to be written in multi-threading programming model which is widely used in existing software.
To support this programming model, we present our initial effort on implementing a language runtime system, CodeFlow, based on the WebAssembly System Interface.
CodeFlow abstracts the underlying architecture and generates code at runtime, relying on just-in-time compilations in WASI.
Looking forward, we envision this programming model is beneficial for future workloads to easily adopt heterogeneous systems.
We are working on extending CodeFlow's compiler and code generation system to support more types of accelerators.
At architecture level, we are implementing CXL accelerators on CXL FPGA to support high-level workloads' requirements, and converting conventional workloads to utilize heterogeneous systems through CodeFlow.

\section{Conclusion}

In this paper, we present CodeFlow, a unified programming model for heterogeneous programming.
CodeFlow employs the emerging CXL interconnection to provide coherent data exchange between heterogeneous devices, and leverages WASI's portable binary format as an intermediate representation between high-level workload and the underlying heterogeneous architecture.
With CodeFlow, application developers can implement workloads as multithreading programs using native language supports, compile it once, and let CodeFlow to execute these threads on heterogeneous architectures.
This programming model decouples the complexity in heterogeneous programming, allowing programmers to focus on designing and optimizing high-level logics, leaving device-specific implementations to system developers, and leveraging a multithreading programming model as a unified intermediate layer.
We envision that with CodeFlow, future workloads can be ported to various heterogeneous systems without rewriting or recompiling code, thus incentivizing the adoption of heterogeneous computing.

% \clearpage

\napaperbib{}

\end{document}